\documentclass[twocolumn,prl,superscriptaddress,preprintnumbers,amsmath,amssymb]{revtex4-1}

\usepackage[paperwidth=210mm,paperheight=297mm,centering,hmargin={1.8cm,1.3cm}, vmargin={1.7cm,2.1cm}]{geometry}
\usepackage[pdftex]{graphicx}
\usepackage{epstopdf}
\usepackage{graphicx}
\usepackage{dcolumn}
\usepackage{bm}
\usepackage{amssymb}
\usepackage{wasysym}
\usepackage{amsmath} 
\usepackage{amsthm}
\usepackage{bibtopic}
\usepackage{sidecap}
\usepackage[rflt]{floatflt}
\usepackage{float}

\begin{document}

\def\Ef{$E_{\rm F}$}
\def\Eb{$E_{\rm B}$}
\def\Efmath{E_{\rm F}}
\def\Ed{$E_{\rm D}$}
\def\Tc{$T_{\rm C}$}
\def\kpara{{\bf k}$_\parallel$}
\def\kparamath{{\bf k}_\parallel}
\def\kperp{{\bf k}$_\perp$}
\def\Gbar{$\overline{\Gamma}$}
\def\Kbar{$\overline{K}$}
\def\Mbar{$\overline{M}$}
\def\BiTe{Bi$_2$Te$_3$}
\def\BiSe{Bi$_2$Se$_3$}
\def\SbTe{Sb$_2$Te$_3$}
\def\Ed{$E_{\rm D}$}
\def\invA{\AA$^{-1}$}

\title{Sub-picosecond spin dynamics of excited states in the topological insulator Bi$_2$Te$_3$}

\author{J. S\'anchez-Barriga}
\email[Corresponding author. E-mail address: ] {jaime.sanchez-barriga@helmholtz-berlin.de.}
\affiliation{Helmholtz-Zentrum Berlin f\"ur Materialien und Energie, Albert-Einstein-Str. 15, 12489 Berlin, Germany}
\author{M. Battiato}
\affiliation{Institute of Solid State Physics, Vienna University of Technology,  Vienna A-1040, Austria} 
\author{M. Krivenkov}
\affiliation{Helmholtz-Zentrum Berlin f\"ur Materialien und Energie, Albert-Einstein-Str. 15, 12489 Berlin, Germany}
\author{E. Golias}
\affiliation{Helmholtz-Zentrum Berlin f\"ur Materialien und Energie, Albert-Einstein-Str. 15, 12489 Berlin, Germany}
\author{A. Varykhalov}
\affiliation{Helmholtz-Zentrum Berlin f\"ur Materialien und Energie, Albert-Einstein-Str. 15, 12489 Berlin, Germany}
\author{A. Romualdi}
\affiliation{Helmholtz-Zentrum Berlin f\"ur Materialien und Energie, Albert-Einstein-Str. 15, 12489 Berlin, Germany}
\author{L. V. Yashina}
\affiliation{Department of Chemistry, Moscow State University, Leninskie Gory 1/3, 119991, Moscow, Russia}
\author{J. Min\'ar}
\affiliation{Department Chemie, Ludwig-Maximilians-Universit\"at M\"unchen, Butenandtstr. 5-13, 81377 M\"unchen, Germany}
\affiliation{New Technologies Research Centre, University of West Bohemia, Univerzitni 2732, 306 14 Pilsen, Czech Republic}
\author{O. Kornilov}
\affiliation{Max-Born-Institut, Max-Born-Str. 2A, 12489 Berlin, Germany}
\author{H. Ebert}
\affiliation{Department Chemie, Ludwig-Maximilians-Universit\"at M\"unchen, Butenandtstr. 5-13, 81377 M\"unchen, Germany}
\author{K. Held}
\affiliation{Institute of Solid State Physics, Vienna University of Technology,  Vienna A-1040, Austria}
\author{J. Braun}
\affiliation{Department Chemie, Ludwig-Maximilians-Universit\"at M\"unchen, Butenandtstr. 5-13, 81377 M\"unchen, Germany}

\begin{abstract}
Using time-, spin- and angle-resolved photoemission, we investigate the ultrafast spin dynamics of hot electrons on the surface of the topological insulator Bi$_2$Te$_3$ following optical excitation by fs-infrared pulses. We observe two surface-resonance states above the Fermi level coexisting with a transient population of Dirac fermions that relax in about $\sim$2 ps. One state is located below $\sim$0.4 eV just above the bulk continuum, the other one at $\sim$0.8 eV inside a projected bulk band gap. At the onset of the excitation, both states exhibit a reversed spin texture with respect to that of the transient Dirac bands, in agreement with our one-step photoemission calculations. Our data reveal that the high-energy state undergoes spin relaxation within $\sim$0.5 ps, a process that triggers the subsequent spin dynamics of both the Dirac cone and the low-energy state, which behave as two dynamically-locked electron populations. We discuss the origin of this behavior by comparing the relaxation times observed for electrons with opposite spins to the ones obtained from a microscopic Boltzmann model of ultrafast band cooling introduced into the photoemission calculations. Our results demonstrate that the nonequilibrium surface dynamics is governed by electron-electron rather than electron-phonon scattering, with a characteristic time scale unambiguously determined by the complex spin texture of excited states above the Fermi level. Our findings reveal the critical importance of detecting momentum and energy-resolved spin textures with fs resolution to fully understand the sub-ps dynamics of transient electrons on the surface of topological insulators.

\end{abstract}

\maketitle
\centerline{\bf I. INTRODUCTION}
\vspace{0.06in}

Topological insulators (TIs) are promising materials for future spintronic applications because they behave as bulk insulators and surface conductors simultaneously \cite{Hasan-RMP-2010, Moore-Nature-2010, Fu-PRL-2007}. Their metallic surface hosts Dirac-cone spin-polarized topological surface states (TSSs) that can be used as channels in which to drive pure spin currents or spin-polarized electrical currents on ultrafast time scales \cite{Pesin-NatMat-2012, Gedik-2012-NatNanotech-photocurrents, Kastl-NatComm-2015, Dankert-NanoLett-2015}. While under equilibrium conditions the helical spin texture of TSSs has been widely studied using spin and angle-resolved photoemission (SARPES) \cite{Hsieh-Science-2009,Hsieh-Nature-2009, Jozwiak-PRB-2011, Pan-PRB-2013, Sanchez-Barriga-PRX-2014}, very little is known about the nonequilibrium spin properties of TSSs following optical excitation by intense fs-laser fields. 

In parallel with material efforts to obtain more bulk-insulating samples so that the bulk contribution to the conductivity can be completely suppressed \cite{Ren-PRB-2011,Arakane-NatComm-2012, Kuroda-PRB-2015}, recent time-resolved (tr) experiments without spin resolution so far revealed the critical role of bulk-mediated electron-phonon scattering in the decay process of hot electrons across the linear energy-momentum dispersion of TSSs \cite{Gedik-PRL-2011-Kerr, Sobota-PRL-2012-Bulk-Reservoir, Hajlaoui-2012-Nanolett-bulk, Gedik-2012-PRL-phonons, Crepaldi-2012-PRB-phonons, Luo-NanoLett-2013-increased-e-ph, Hajlaoui-NatComm-2014-e-h-asym, Reimann-2014-PRB-phonons, Sobota-PRL-2014-Oscillations}. These findings indicate promising routes to overcome the problem of the bulk conductivity on ultrafast time scales, for example by generating low-energy excitations using fs-laser pulses, so that only electrons transiently occupying the TSS are excited within the bulk band gap \cite{Gedik-2013-Science-Floquet}. Experiments along this line have brought encouraging results, such as the emergence of photon-dressed Dirac bands establishing the observation of an insulating Floquet phase in TIs \cite{Gedik-2013-Science-Floquet, Gedik-2016-NaturePhys-Floquet}, although the spin properties of such exotic excitations have remained unexplored. These experiments also established that a dynamical gap can be opened at the Dirac node of the TSS due to strong coupling to a fs-laser field generated with circularly-polarized photons \cite{Gedik-2013-Science-Floquet}, in contrast to the case of low-energy excitations induced by ultrashort mid-infrared pulses of linearly-polarized light \cite{Kuroda-PRL-2016}. Although all these observations taken collectively might pave the way for the realization of a transient-anomalous quantum Hall effect on ultrafast time scales without the need of magnetic dopants or applied magnetic fields \cite{Dahlhaus-PRL-2016}, spin-resolved measurements might prove crucial to further clarify it.

\begin{figure*}
\centering
\includegraphics [width=0.99\textwidth]{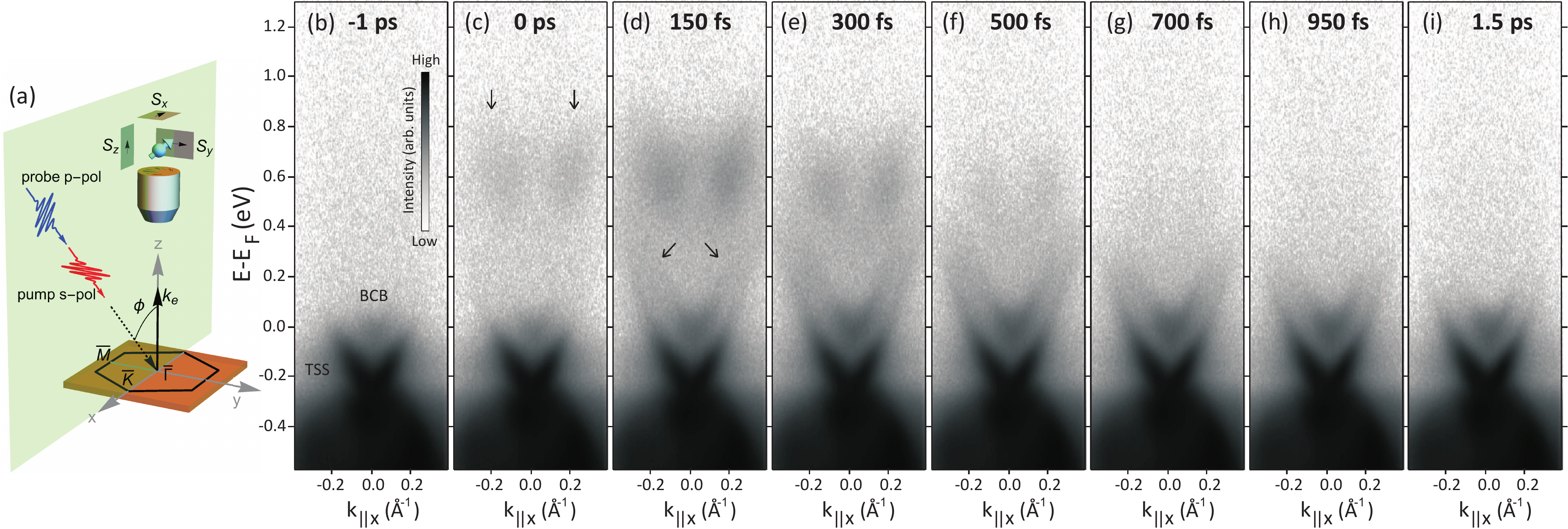}
\caption{(a) Experimental geometry. The light incidence plane is oriented along the \Gbar-\Kbar\ direction of the SBZ, and the pump and probe pulses are linearly $s$- and $p$-polarized, respectively. S$_x$, S$_y$ and S$_z$ denote the corresponding projections of the spin polarization. (b)-(i) Tr-ARPES spectra of Bi$_2$Te$_3$ obtained along \Gbar-\Kbar\ at various pump-probe delays, as indicated on the top of each panel. In (b), the equilibrium band structure is probed before optical excitation. In (c)-(i), the dynamics of excited electrons above the Fermi level is probed following optical excitation. Besides the transiently populated Dirac bands, high-energy states exhibiting faster dynamics are observed up to $\sim$0.9 eV above the Fermi level, while low-energy states appearing as a continuation of the bulk continuum are observed up to $\sim$0.5 eV. The states are highlighted with arrows in (c) and (d), respectively. The spectra were obtained at $300$ K with pump and probe pulses of 1.5 and 6 eV photon energy, respectively.}
\label{Figure1}
\end{figure*}

Equally important, the impact of bulk-to-surface coupling on the ultrafast spin dynamics of TSSs following a high-energy excitation outside the bulk band gap is not yet understood. First experiments on prototypical TIs combining time, spin, energy and momentum resolutions (tr-SARPES) indicate that, for sufficiently $n$-doped samples, bulk and surface bands behave as independent relaxation channels for the electron spin \cite{Cacho-Surface-resonances-PRL-2014}, while for $p$-doped samples, as two-coupled channels where the spin-dependent scattering rates for bulk electrons are about one order of magnitude higher than for surface electrons \cite{Sanchez-Barriga-PRB-2016}.
 
As a consequence of the weak surface scattering these findings enabled, on the one hand, the observation of spin-polarized electrical currents originating from TSSs using circularly-polarized light in the time domain \cite{Kastl-NatComm-2015, Sanchez-Barriga-PRB-2016, Boschini-SciRep-2016}. On the other hand, the use of linearly-polarized fs-infrared pulses enabled the suppression of the charge current on the surface \cite{Gedik-2012-NatNanotech-photocurrents, Gedik-PRL-2011-Kerr}, presumably giving rise to pure ultrafast spin currents evolving on a different time scale \cite{Gedik-PRL-2011-Kerr}, so that the transient Dirac cone can be dynamically populated by the same number of excited electrons moving in opposite directions with opposite spins \cite{Sobota-PRL-2012-Bulk-Reservoir, Kuroda-PRL-2016}. However, mainly due to the lack of spin resolution in previous tr-ARPES experiments, it has not been investigated so far whether or not the scattering mechanisms underlying the ultrafast electron relaxation of TSSs strongly depend on the spin of the excited states, or how spin-selection rules affect the sub-ps dynamics of TIs. The answer to these questions might in fact prove crucial to further understand the role of the electron spin in the relaxation pathways of the generated spin currents, and whether or not such currents directly follow the time scale of the Dirac-cone electron-momentum relaxation. These aspects are especially important as, in the presence of strong spin-orbit coupling, the ultrafast processes of electron-momentum relaxation might be strongly constrained by spin selection at the surface \cite{Pascual-PRL-2004, Roushan-Nature-2009} or other scattering events that involve the electron spin such as the ones arising from the effective coupling between bulk and surface-state electrons \cite{Wang-PRL-2016, Sim-PRB-2014}.

To investigate these issues, in the present work we perform tr-SARPES measurements on the prototypical TI Bi$_2$Te$_3$ following optical excitation by fs-infrared pulses of linearly-polarized light. We observe spin-polarized electron excitations up to $\sim$1 eV above the Fermi level which decay through an avalanche of hot electrons within less than $\sim$2 ps. We explore the nonequilibrium ultrafast spin dynamics of the photoexcited states and disentangle the surface contributions to the transient spin polarization. To understand the underlying mechanisms that trigger the observed behavior of the nonequilibrium spin populations, we perform one-step photoemission calculations in the framework of a microscopic model that includes all types of electron scatterings on sub-ps timescales. Our results demonstrate that the characteristic time scale for electron relaxation is governed by electron-electron scattering processes that are ultimately determined 
by the complex spin texture of excited states above the Fermi level. Our findings are of critical importance for understanding the role of the electron spin in the ultrafast dynamics of Dirac fermions in TIs.
\begin{figure*}
\centering
\includegraphics [width=0.75\textwidth]{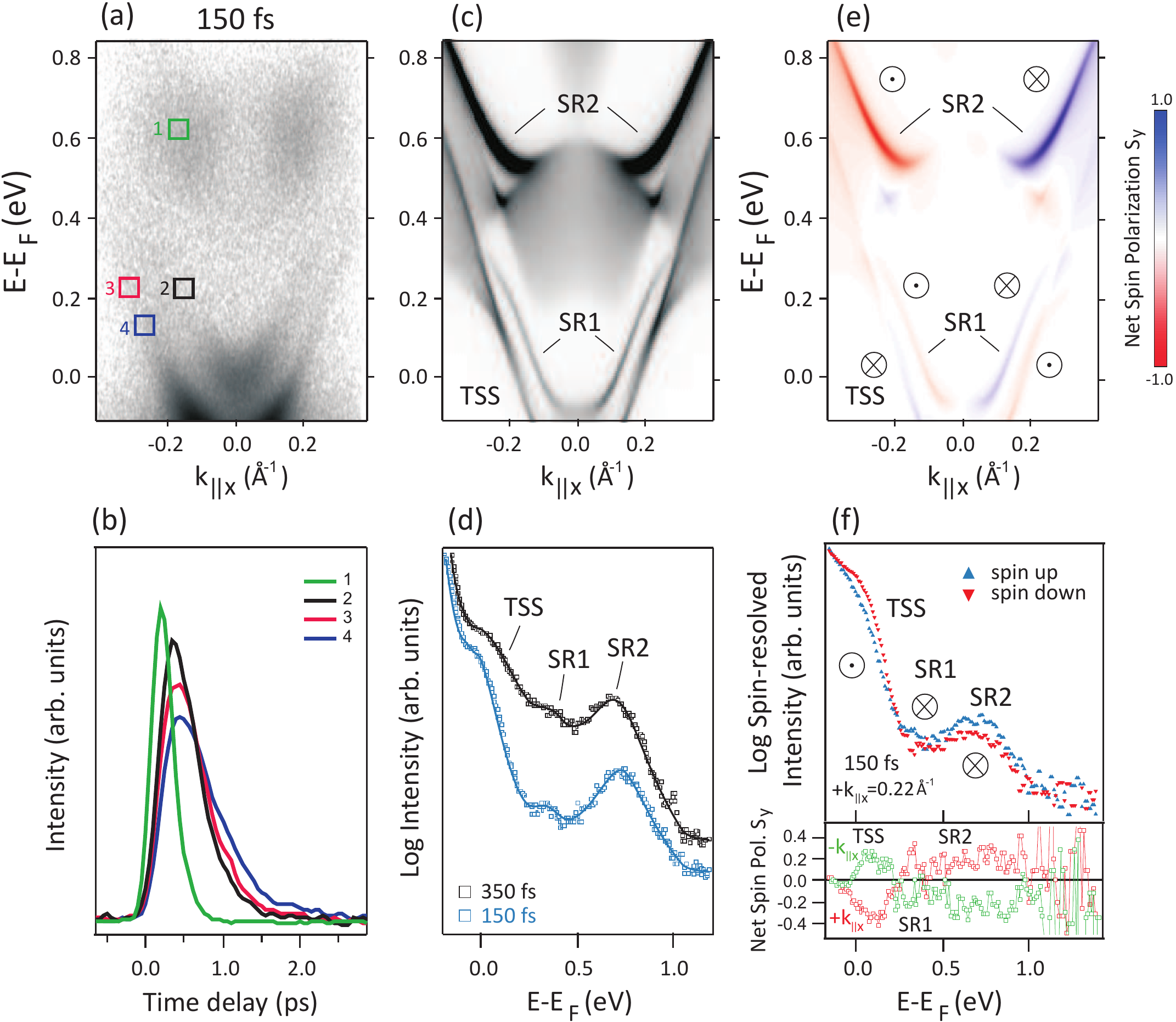}
\caption{(a) Tr-ARPES spectrum of Bi$_2$Te$_3$ obtained 150 fs after optical excitation. (b) Corresponding tr-ARPES intensity integrated within small energy-momentum windows [labeled from 1 to 4 among different states in (a)]. (c) One-step model photoemission calculations under equilibrium conditions, revealing contributions from the TSS and trivial surface resonances (denoted as SR$1$ and SR$2$). (d) Experimental EDCs extracted at $\sim$0.22\invA\ and different time delays, containing contributions from the TSS, SR1 and SR2 states at different energies. (e) Calculations of the spin-polarized electronic structure corresponding to (c). The red-blue color intensity is proportional to the magnitude of the tangential component of the spin polarization S$_y$, and its orientation perpendicular to momentum is indicated by the symbols $\odot$ and $\otimes$, which are vectors pointing out and into the paper plane, respectively. (f) Top panel: Spin-resolved EDCs measured 150 fs after optical excitation at the same wave vector as the EDCs shown in (d). The blue (red) curves are tangential spin up (down) EDCs, and the direction of the spin polarization component S$_y$ of the TSS, SR$1$ and SR$2$ states is denoted in accordance to (e). Bottom panel: Corresponding net spin polarization S$_y$, which reverses for SR$1$ and SR$2$ states with respect to the TSS as well as at opposite wave vectors (red and green colors, respectively.)}
\label{Figure2}
\end{figure*}

\vspace{0.15in}
\centerline{\bf II. METHODS}
\vspace{0.03in}

Experiments were performed at room temperature under ultrahigh vacuum conditions with a base pressure below $1\cdot10^{-10}$ mbar. Photoelectrons and the three projections of their spin polarization were detected with a Scienta R4000 hemispherical analyzer coupled to a Mott-type spin detector operated at 26 kV. Bi$_2$Te$_3$ single crystals were grown by the Bridgman method and cleaved {\it in situ}. The first (1.5 eV) and fourth (6 eV) harmonics of a homemade fs-laser system coupled to an ultrafast amplifier operating at 100 kHz repetition rate were used as pump and probe pulses, respectively. The time resolution was $\sim$200 fs, and the pump fluence $\sim$100 $\mu$J/cm$^{2}$. The angular and energy resolutions of tr-ARPES measurements were 0.3$^{\circ}$ and 30 meV, respectively. Resolutions of tr-SARPES measurements were 0.75$^{\circ}$ (angular) and 80 meV (energy). We used linearly $s$- and $p$-polarized pump and probe pulses unless otherwise specified, respectively, incident on the sample under an angle of $\phi=45^{\circ}$ following the experimental geometry shown in Fig. 1(a). Photoemission calculations are based on multiple scattering theory within the one-step model of photoemission in its spin-density matrix formulation, including wave-vector, spin and energy-dependent transition matrix elements \cite{Hopkinson-CPC-1980, Braun-theory-96}. We use a fully-relativistic version that is part of the spin polarized relativistic Korringa-Kohn-Rostoker (SPR-KKR) program package, with spin-orbit coupling included self-consistently  \cite{Ebert-SPRKKR-2011, Ebert-SPRKKR-2012}. The time-dependent photoemission calculations are performed within the Boltzmann approach \cite{Fatti-PRB-2000, Rethfeld-PRB-2002, Mueller-PRB-2013}, using the SPR-KKR bands as well as the results of the one-step model calculations as an input. The dynamical calculations take into account all possible spin-dependent electron transitions in energy-momentum space as well as electron-phonon scatterings in a quantitative way (for more details please see discussion below and Supplemental information \cite{Supplemental-Material}).

\vspace{0.1in}
\centerline{\bf III. RESULTS AND DISCUSSION}
\vspace{0.03in}
{\bf A. Sub-ps electron dynamics and one-step model photoemission calculations}
\vspace{0.03in}

To understand the nonequilibrium dynamics of the excited states above the Fermi level, in Figs. 1(b)-1(i) we show tr-ARPES measurements recorded at various pump-probe delays near the \Gbar\ point of the Bi$_2$Te$_3$\ surface Brillouin zone (SBZ). Before optical excitation and at negative time delays [Fig. 1(b)], we probe the electronic band structure in equilibrium, as evidenced by the lack of photoemission intensity above the Fermi level. Due to the $n$-doping, the minimum of the bulk-conduction band (BCB) and part of the TSS are observed below the Fermi level, with the Dirac node located at an energy of $E_D\sim$-0.29 eV. At the onset of the optical excitation [Fig. 1(c)], an initial population of higher-energy states located at $\sim$0.7 eV above the Fermi level is clearly observed [marked with arrows in Fig. 1(c)]. During the subsequent dynamics [Figs. 1(d)-1(f)], these higher-energy states rapidly decay within less than $\sim$1 ps. Besides the transiently populated Dirac bands, other states exhibiting a similar energy-momentum dispersion are observed up to $\sim$0.4 eV. These lower-energy states [marked with arrows in Fig. 1(d)] appear as a continuation of the bulk continuum and above the BCB top, which is located in the immediate vicinity of the Fermi level. Remarkably, in Figs. 1(g)-1(i) we observe that the transient Dirac bands and the lower-energy states decay synchronously, a process which only occurs once the higher-energy states have completely relaxed. The latter behavior strongly indicates that there is an effective electron transfer from the higher-energy states into both the transient Dirac bands and the lower-energy states. 

Such an electron transfer process as well as the synchronous electron decay can be further visualized in Figs. 2(a) and 2(b), where we analyze the integrated tr-ARPES intensity [Fig. 2(b)] within small energy-momentum windows [labeled from 1 to 4 among the different states in Fig. 2(a)]. Very clearly, while the tr-ARPES intensity of the higher-energy states in Fig. 2(b) decays rapidly (window 1), the intensities of the lower-energy states (window 2) and TSS bands (windows 3 and 4) exhibit very similar dynamics, so that electrons with the same energy but different momenta (windows 2 and 3) display similar relaxation times. This process is preceded by a delayed electron filling of the TSS bands and the lower-energy states, initially causing rise times in their intensities (windows 2-4) that correlate well with the overall decay of the tr-ARPES intensity from the higher-energy states. We also point out that the relaxation time scales are weakly dependent on the pump fluence (for more details please see Supplemental information \cite{Supplemental-Material}). 

To further understand the nature of the higher and lower-energy states, in Fig. 2(c) we show the results of one-step model photoemission calculations performed under equilibrium conditions using linearly $p$-polarized 6 eV photons. Differently from previous studies on prototypical TIs \cite{Hajlaoui-2012-Nanolett-bulk, Hajlaoui-NatComm-2014-e-h-asym, Cacho-Surface-resonances-PRL-2014}, our calculations reveal that both states are surface-state-like features containing a rather low bulk contribution, allowing us to unambiguously identify them as topologically trivial surface resonances \cite{Echenique-JPC-1978, McRae-RMP-1979} (labeled as SR1 and SR2) dispersing near the border of two different bulk-projected band gaps (see Supplemental information for details \cite{Supplemental-Material}). In particular, the higher-energy state (SR2), is located inside and at the border of a bulk-projected band gap appearing at twice higher energy than the main gap of the volume, while the low-energy state (SR1), is located outside and at the border of the main bulk-band gap. In the experiment, the relative contribution of the different states can also be resolved simultaneously by exploiting the time and momentum resolutions through energy-distribution curves (EDCs) extracted at off-normal wave vectors [Fig. 2(d)]. Overall, our theoretical results lead to good qualitative agreement with the experiment concerning both the relative energy positions of the observed states as well as their energy-momentum dispersions. 
\begin{figure*}[t]
\centering
\includegraphics[width=0.99\textwidth]{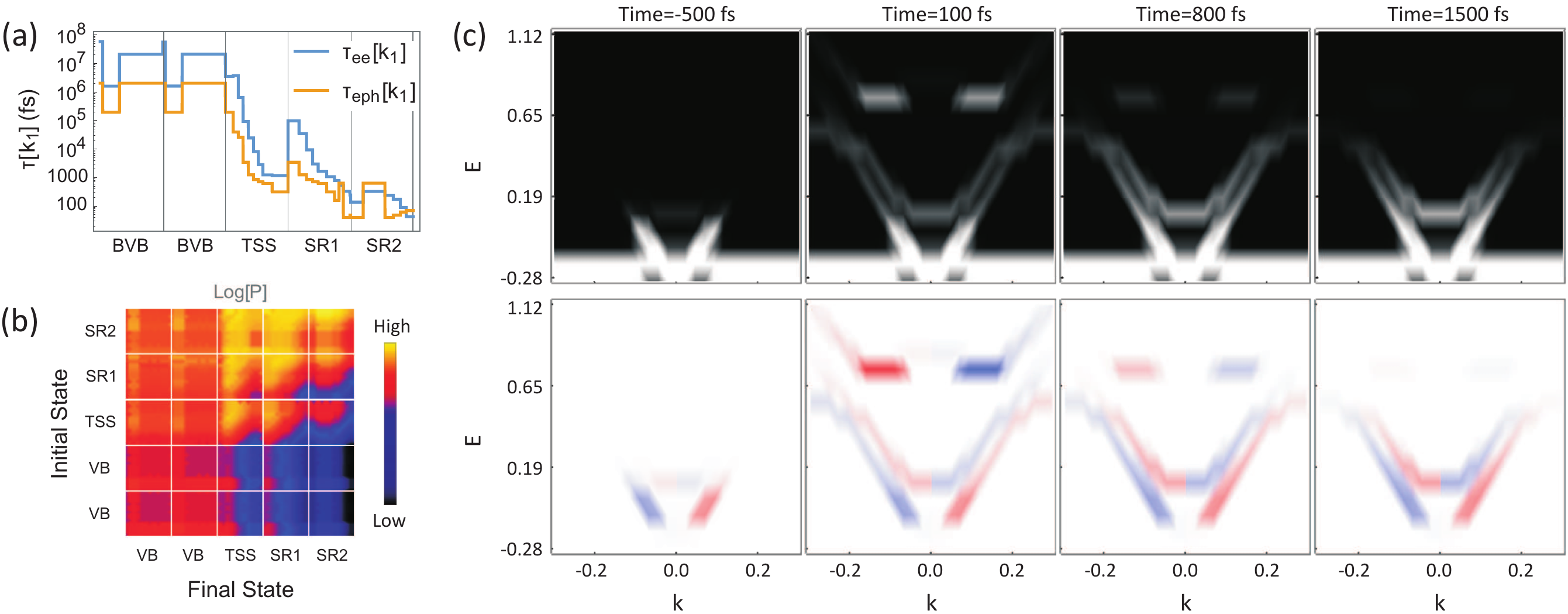} 
\caption{(a) Electron-electron and electron-phonon lifetimes deduced from the scattering probabilities assuming an electron and phonon population at $300$ K. The $x$-axis represents the momentum k and is split among the bands considered in the dynamics calculation. The k value for all the bands goes from $0$ to $0.3$ \invA\ from left to right. (b) Probability of an electron in a given band with a given momentum k to scatter and end up in another band with different momentum. Both axis represent the momentum k and are split among the bands. The k value for all the bands goes from $0$ to $0.3$ \invA\ from left to right and from top to bottom. (c) Calculated snapshots of the momentum and energy-resolved total electronic population (top row) and tangential component of the spin polarization S$_y$ (bottom row) at different time delays. Note that the calculated spin polarization is constant in time. Only at delays in which the total intensity of a given state is very small, the spin polarization falls due to an unpolarized constant background that was introduced in the calculation to avoid zero by zero divisions.}
\label{Figure3}
\end{figure*}

\vspace{0.06in}
{\bf B. Transient spin polarization of excited states above the Fermi level}
\vspace{0.06in}

Most interestingly, if we examine the tangential component of the spin polarization S$_y$ in our calculations [Fig. 2(e)], which is perpendicular to the electron momentum, we find that besides the expected helical spin texture from TSSs \cite{Hsieh-Nature-2009}, the surface-resonance states exhibit a reversed spin texture with respect to that of the Dirac bands. This prediction is fundamentally different from what is expected for pure bulk states, as even in the presence of strong spin-orbit coupling, bulk bands in Bi$_2$Te$_3$ are essentially unpolarized. Therefore, to investigate the spin orientation of the states above the Fermi level experimentally, we perform tr-SARPES measurements at $k_{\parallel}\sim0.22 $\invA\ and a fixed time delay of 150 fs following optical excitation [see Fig. 2(f)]. Here, we focus on the previously unmeasured transient spin polarization of Bi$_2$Te$_3$ above the Fermi level. In accordance with the EDCs shown in Fig. 2(d), the results of Fig. 2(f) contain a contribution from all different states. In addition to the intensities of the tangential spin up and spin down populations, which are resolved independently [top panel in Fig. 2(f)], we show the corresponding S$_y$ component of the spin polarization, which reverses at opposite wave vectors [bottom panel in Fig. 2(f)]. We note that while the measured S$_x$ component of the spin polarization is zero, we find a small out-of-plane spin component S$_z$ in both experiment and calculations (see supplemental information \cite{Supplemental-Material}). Moreover, using $s$-polarized probe pulses instead causes a reversal of the observed spin polarization \cite{Supplemental-Material}. This result is consistent with the expected spin-orbital texture of the excited states as derived from our calculations, implying that with $p$-polarized probe pulses we are sensitive to the spin projections of p$_y$ and p$_z$ orbitals \cite{Zhu-PRL-2013}. On the other hand, changing the pump polarization does not influence the observed spin texture \cite{Supplemental-Material}, indicating that the measured spin polarization originates from ground-state properties rather than from pump-induced spin-dependent matrix elements \cite{Optical-Orientation-1984, Supplemental-Material}. 

Overall, the observed spin polarization is qualitatively similar to the results of our theoretical calculations. If we compare the experimental and theoretical values of the spin polarization for different states, we find that for SR2 states the calculation overestimates their absolute spin polarization by about $\sim$ 20\%, while for the TSS and SR1 states, the spin polarization is underestimated by about $\sim$10\%. We attribute this disagreement to the fact  that we use the atomic sphere approximation within the DFT-LDA approach \cite{Braun-theory-96}, which differently from GW self-energy corrections  \cite{Foester-PRB-2015}, introduces variations in the calculated spin polarization. Despite these small deviations, we find excellent agreement between theory and experiment concerning the spin structure of the states, implying that what we have indeed observed in our tr-ARPES experiments is a dynamical transfer of electrons between states with opposite spin textures. At the same time, this observation indicates that the overall dynamics is a consequence of rather complex scattering processes, possibly requiring large momentum transfers. The reason is that decay of electrons between states with opposite spins is quantum-mechanically forbidden unless it involves a spin flip, which as long as time-reversal symmetry is preserved \cite{Hsieh-PRL-2009}, is the less probable process.

Another important observation is the symmetric tr-ARPES intensity distribution of the different bands at opposite wave vectors, as it implies that there is a suppression of the transient charge current on the surface \cite{Sanchez-Barriga-PRB-2016, Kuroda-PRL-2016, Guede-PRB-2015}. Thus, our tr-SARPES measurements additionally demonstrate, on the one hand, that linearly-polarized infrared pulses induce pure ultrafast spin currents originating from Dirac fermions and, on the other hand, that  on fs time scales such currents coexist with trivial spin currents of reversed direction of the spin polarization. This finding implies that combining time, spin, energy and momentum resolutions in a single experiment is crucial to unambiguously identify the nature of the generated currents. Our results also provide further insight on which electronic states could be responsible for the reflectivity dynamics and complex Kerr signal observed in recent state-of-the-art experiments on the same material \cite{Boschini-SciRep-2016}.

\vspace{0.06in}
{\bf C. Time-resolved photoemission calculations and spin-dependent scattering mechanisms}
\vspace{0.06in}

Questions arise on how the alternating spin polarization of the bands affects the dynamics, and which are the scattering mechanisms underlying the ultrafast electron relaxation. To address this we performed calculations of the dynamics within the Boltzmann approach \cite{Fatti-PRB-2000, Rethfeld-PRB-2002, Mueller-PRB-2013}, and analyzed the contribution from electron-electron and electron-phonon scatterings in a quantitative way. The probability density of an electron with momentum $\mathbf{k}_1$ to scatter with an electron at momentum $\mathbf{q}_1$ and make a transition to final states $\mathbf{k}_2$ and $\mathbf{q}_2$ is given by the transition matrix elements $|\left< \mathbf{k}_1 \mathbf{q}_1 \right|  \hat{H}_{\mbox{\tiny{e-e}} } \left| \mathbf{k}_2 \mathbf{q}_2 \right>|^2$ with the constrain that the total energy and momentum are preserved upon the scattering, giving rise to the two terms $\delta\left(\mathcal{E}\left(\mathbf{k}_1\right) +\mathcal{E}\left(\mathbf{q}_1\right) -\mathcal{E}\left(\mathbf{k}_2\right) -\mathcal{E}\left(\mathbf{q}_2\right)  \right)  $  and $\delta\left(\mathbf{k}_1 +\mathbf{q}_1 -\mathbf{k}_2 -\mathbf{q}_2  \right) $, where $\delta(\cdot)$ is the Dirac delta function, $\mathcal{E}\left(\mathbf{k}\right)$ is the band dispersion, and, for brevity, the band indexes have been dropped. In our calculations we assume the transition matrix element to be weakly dependent on the spatial part of the wave function, while keeping the spin dependence in an approximate way as the overlap of the initial and the final spin states
\begin{align} 
	|\left< \mathbf{k}_1 \mathbf{q}_1 \right|  \hat{H}_{\mbox{\tiny{e-e}} } \left| \mathbf{k}_2 \mathbf{q}_2 \right>|^2 \approx &\;\;\mathcal{A} \left( \frac{1+\overline{s}_{\mathbf{k}_1}\cdot \overline{s}_{\mathbf{k}_2}}{2}  \frac{1+\overline{s}_{\mathbf{q}_1}\cdot \overline{s}_{\mathbf{q}_2}}{2} + \right. \nonumber \\
	&\left. \frac{1+\overline{s}_{\mathbf{k}_1}\cdot \overline{s}_{\mathbf{k}_2} }{2} \frac{1+\overline{s}_{\mathbf{q}_1}\cdot \overline{s}_{\mathbf{q}_2}}{2} \right)
\end{align}
where $\mathcal{A}$ is a constant, $\overline{s}_{\mathbf{k}_1} = \left< \mathbf{k}_1  \right|  \hat{s} \left| \mathbf{k}_1 \right>/|\left< \mathbf{k}_1  \right|  \hat{s} \left| \mathbf{k}_1 \right>|$ is the unit vector oriented along the spin expectation value on the state $\mathbf{k}_1$ and $\hat{s} $ is the vector spin operator.

To study the ultrafast relaxation of the transient spin populations only depending on the distance from \Gbar\ in a simplified way, we approximate the SBZ as circular and assume the excitation to have circular symmetry within the SBZ \cite{Supplemental-Material}. We will see that the mentioned approximation captures extremely well the relaxation times of spin up and spin down electrons within different bands. The reason is that the overall dynamics of the excited states tends to quickly converge towards SBZ center, while the thermalization of the bands as well as the exchange of electrons between bands is a strongly spin-dependent process. Hence, we need to construct the probability $\mathcal{P}(k_1,q_1,k_2,q_2)$ of an electron with momentum $\mathbf{k}_1$, such as $\left|\mathbf{k}_1\right|=k_1$, to scatter with an electron at any $\mathbf{q}_1$ such as $\left|\mathbf{q}_1\right|=q_1$ and make a transition to any final states $\mathbf{k}_2$ and $\mathbf{q}_2$, such as $\left|\mathbf{k}_2\right|=k_2$ and $\left|\mathbf{q}_2\right|=q_2$. The expression for $\mathcal{P}(k_1,q_1,k_2,q_2)$ can be obtained from geometrical arguments but it is lengthy and reported only in the Supplementary Material \cite{Supplemental-Material}. The change in time of the total population $n(k,t)$ at a given $\mathbf{k}$ such as $\left|\mathbf{k}\right|=k$ due to electron-electron scatterings is given by:
\begin{align}
	&\left(\frac{\partial n(k,t)}{\partial t}\right)_{\mbox{e-e}} = \int dk_1 dq_1 dq_2 \\
	&\left(\mathcal{P}(k_1,q_1,k,q_2) n(k_1,t) n(q_1,t)\left( 1\!-\!n(k,t)\right)\left( 1\!-\!n(q_2,t)\right) \right.  \nonumber\\
	&\left.- \mathcal{P}(k,q_1,k_1,q_2) n(k,t) n(q_1,t)\left( 1\!-\!n(k_1,t)\right)\left( 1\!-\!n(q_2,t)\right) \right. \nonumber 
\end{align}

We treat the electron-phonon scatterings in a similar, but more simplified way. The high number of phonon states within a small energy window (compared to the electronic excitations and the distances between the electronic bands), allows us to neglect here the geometrical constraints due to momentum conservation and yields 
\begin{widetext}
\begin{equation}
\begin{split}
\left(\frac{\partial n(k,t)}{\partial t}\right)_{\mbox{ph}} = \mathcal{B} \int dk_1& \left[  n(k_1,t)\left( 1\!-\!n(k,t)\right) \left(f_{\tiny{\mbox{BE}}}(\delta\mathcal{E}) \rho(\delta\mathcal{E})+(1+f_{\tiny{\mbox{BE}}}(-\delta\mathcal{E})) \rho(-\delta\mathcal{E}) \right) \right. \\
	&\left. - n(k,t)\left( 1\!-\!n(k_1,t)\right) \left(f_{\tiny{\mbox{BE}}}(-\delta\mathcal{E}) \rho(-\delta\mathcal{E})+(1+f_{\tiny{\mbox{BE}}}(\delta\mathcal{E})) \rho(\delta\mathcal{E}) \right) \right]
\end{split}
\end{equation}
\end{widetext}
where $\delta\mathcal{E}=\mathcal{E}(k)-\mathcal{E}(k_1)$, $\mathcal{B}$ is a constant that represents the electron-phonon coupling, $\rho$ is the phonon density of states, and $f_{\tiny{\mbox{BE}}}$ is the Bose-Einstein distribution at the corresponding phononic temperature.  

We turn now to the actual calculations, the results of which are shown in Figs. 3(a)-3(c). The phononic temperature is room temperature, and the density of states of the phonons is constant up the Debye frequency ($\hbar\omega_D$=15 meV) \cite{Hellman-PRB-2015}. The model for the dynamics has therefore only two parameters, $\mathcal{A}$ the electron-electron and $\mathcal{B}$ the electron-phonon coupling. Our calculations below reveal that both types of scatterings lead to completely different dynamical behaviors associated to the specific relaxation of the different bands. This allows the unambiguous estimation of the two parameters $\mathcal{A}$= 0.006 nm$^3$/fs and $\mathcal{B}\rho(\mathcal{E})$ = 0.333 $\Theta(\hbar\omega_D - \mathcal{E})\Theta(\mathcal{E})$ nm/fs, where $\Theta$ is the Heaviside step function.

At early times after optical excitation the higher-energy surface resonance (SR2) is heavily populated by the pump pulse [see Fig. 3(c)]. Electron-phonon scatterings contribute in removing energy from the electronic system, however this type of scatterings turn out to be inefficient in transferring SR2 electrons to any lower-energy bands. This can be seen in Fig. 3(a), where we show the comparison of electron-electron and electron-phonon scattering lifetimes computed at $300$ K for different bands. In fact, we can see that the phonon-mediated energy transfer mechanism from SR2 states has extremely low efficiency, and every scattering event removes very small amount of energy. Conversely, a single electron-electron scattering considerably lowers the energy of the scattered electron. A transient SR2 electron can directly scatter with an electron below the Fermi energy. The most probable outcome of such a process will be both electrons occupying states of the transient Dirac bands or the lower-energy surface resonance (SR1). To illustrate this in more detail, in Fig. 3(b) we show the probability that an electron in a state $k_1$ makes a transition to a state $q_1$ assuming that all the other electrons are distributed at 330 K. Two main scatterings mechanisms are responsible for the decay of the photoemission intensity of the high-energy states: low-$k$-transfer scatterings to the lower-energy surface resonance (SR1) and high-$k$-transfer ones directly to the Dirac cone. Moreover, their probabilities are further increased by the presence of numerous surface-like valence band (VB) states below the Dirac node \cite{Supplemental-Material}, which act as scatterers.

The transient populations of the lower-energy surface resonance (SR1) and the TSS bands, instead, display much longer electron-electron scattering lifetimes, as seen in Fig. 3(a). Electrons in the TSS and SR1 states can reduce their energy in two ways: by changing bands or by moving down within the same band. A reduction of energy within the same band requires, due to the spin selection, a transfer of linear momentum which has to be small and directed towards the center of the SBZ. This reduces the calculated possible transitions for the second electron, compatible with energy, momentum and spin selection rules. In particular, due to this restriction, our calculation shows that the most probable transition for the second electron can only be from a state in the VB to either the TSS or the SR1 bands on the opposite side of the SBZ. However, these transitions are only possible if the energy transfer is higher than the binding energy of the original VB state. Conversely, scatterings that cause a transition from SR1 to TSS or vice versa require a large $k$ transfers. Due to the almost conical nature of the bands, it can be shown that the most probable transition only occurs when the final (initial) energy of the second electron is the same as the initial (final) energy of the first. It is evident that these type of transitions are not effective in redistributing energy. Thus, as shown by our calculations in Fig. 3(a), the equilibration of the TSS and SR1 states can only be accounted for if additionally driven by electron-phonon scatterings. The latter have less strict selection rules. Nevertheless, the limitations due to the complex spin texture of excited states are still that high-$k$-transfer (low-$k$-transfer) electron transitions are forbidden if the initial and final band is the same (different), and that the total change in energy achievable is small due to the small energy of the phonons. However, the energy is efficiently taken out of the electronic system, and this process becomes relevant in the immediate vicinity of the Fermi level once electron relaxation due to electron-electron scattering at higher energies has evolved in time scale slower than expected due to the alternating spin texture of excited states. Nevertheless, note that at low energy the presence of several phonon bands up to 15 meV ensures that there will be always available phonon transitions for any $k$-transfer necessary to accomplish the specific transition.
\begin{figure}[!tbp]
\centering
\includegraphics[width=0.42\textwidth]{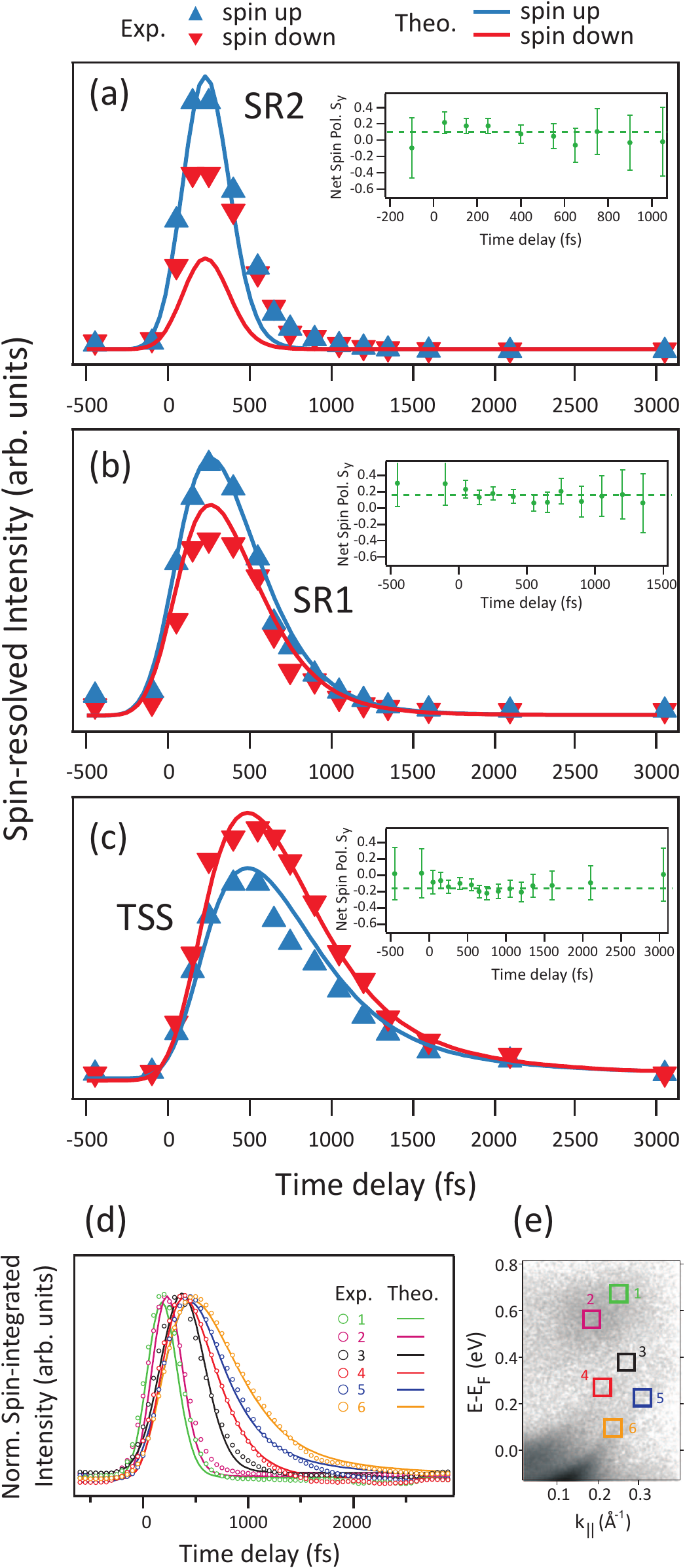} 
\caption{Detailed comparison between tr-SARPES experiments and dynamical calculations. (a)-(c) Dynamics of spin up and spin down electrons within different bands, obtained at $k_{\parallel}\sim0.22 $\invA\ and at the energies of the (a) SR$2$, (b) SR$1$ and (c) TSS bands. Blue and red triangles (solid lines) denote experimental (calculated) spectra for opposite spin. The corresponding spin polarization S$_y$ measured as a function of pump-probe delay is shown in the insets of (a)-(c). Green dashed lines are guides to the eye. The error bars are given by $\Delta S=\sqrt{1/NS^2}$, where $S$=0.2 is the Sherman function and $N$ the corresponding spin-integrated counts. (d) Normalized spin-integrated intensities obtained within the small energy-momentum windows shown in (e) as a function of pump-probe delay. Symbols denote the experimental spectra, solid lines are the corresponding calculations.}
\label{Figure4}
\end{figure}

\vspace{0.01in}
{\bf D. Momentum- and energy-resolved spin dynamics: Comparison theory-experiment}
\vspace{0.01in}

If we compare in detail the results of our dynamical calculations to the experiment (see Fig. 4), we find good quantitative agreement concerning both the dynamics of tangential spin up and spin down electrons within different bands [Figs. 4(a)-4(c)] as well as the decay of the spin-integrated intensities [Fig. 4(d)] within various energy-momentum windows distributed over different states [Fig. 4(e)]. Specifically, the measurements shown in Figs. 4(a)-4(c) were taken by integrating the spin-resolved intensities as a function of time delay at a fixed wave vector $k_{\parallel}\sim0.22 $\invA\ and at the corresponding energies of the SR2 [Fig. 4(a)], SR1 [Fig. 4(b)] and TSS bands [Fig. 4(c)], in accordance with the results of Fig. 2(f). Similarly, the theoretical curves were extracted from small energy-momentum boxes superimposed on top of the calculated band dispersions of Fig. 3(c) at the corresponding energies and wave vectors. Accordingly, in each inset of Figs. 4(a)-4(c) we show the measured tangential components of the spin polarization S$_y$ as a function of pump-probe delay. It is seen that the measured spin polarizations stay constant in time within the experimental error bars, a result that is consistent with our calculations of Fig. 3(c) and allows us to exclude that the spin texture of the transient electronic states is dynamically modulated by the pump excitation itself, despite the nonequilibrium condition. We have also probed the dynamics of other components of the spin polarization, and obtained similar behavior \cite{Supplemental-Material}. Thus, the decay times of spin up ($\tau^{\uparrow}$) and spin down ($\tau^{\downarrow}$) electrons in Fig. 4 are representative for the three-dimensional spin dynamics. 

In particular, by fitting a single-exponential decay function to the data, for SR2 states we obtain experimental (theoretical) relaxation times of $\tau^{\uparrow}$= 192.5 (134.6) $\pm$ 41.3 (6.4) fs and $\tau^{\downarrow}$= 210.1 (135.3) $\pm$ 43.9 (5.9) fs, while for SR1 states $\tau^{\uparrow}$= 291.7 (311.5) $\pm$ 20.1 (3.5) fs and $\tau^{\downarrow}$= 309.5 (311.6) $\pm$ 24.9 (3.6) fs. Likewise, for the TSS we obtain $\tau^{\uparrow}$= 594.6 (583.1) $\pm$ 58.4 (15.4) fs and $\tau^{\downarrow}$= 577.2 (582.8) $\pm$ 61.8 (15.4) fs. The similarity between the relaxation times of electrons with opposite spins for each individual state, as well as the quantitative agreement between experiment and theory, implies that the overall relaxation process of excited electrons is driven by surface-dominated dynamics. We emphasize that completely different relaxation times for electrons with opposite spins would be instead expected in the case of a simultaneous contribution from two independently-thermalizing bulk and surface populations relaxing on different time scales, especially if the transient electronic temperatures of bulk and surface states differ substantially \cite{Sobota-PRL-2012-Bulk-Reservoir, Cacho-Surface-resonances-PRL-2014}. This scenario would be the consequence of a complex interplay between bulk and surface dynamics \cite{Cacho-Surface-resonances-PRL-2014}, so that one spin component is dominated by the time scale of the surface contribution while the other one by the larger relaxation times of bulk states. In fact, previous works without spin resolution have put forward bulk-assisted dynamics as one of the relevant mechanisms underlying the relaxation times of hot electrons in TIs \cite{Gedik-PRL-2011-Kerr, Sobota-PRL-2012-Bulk-Reservoir,  Gedik-2012-PRL-phonons, Hajlaoui-2012-Nanolett-bulk}. However, our theoretical and experimental findings show that the nonequilibrium dynamics of a prototypical TI such as Bi$_2$Te$_3$ is fully explained by the surface contribution, a conclusion that can only be realized by directly probing the spin texture of excited states above the Fermi level with time, energy and momentum resolution. In addition, our results demonstrate that the alternating spin texture of the excited states completely determines the relevant scattering channels for electron relaxation.
\begin{figure}[!tbp]
\centering
\includegraphics[width=0.35\textwidth]{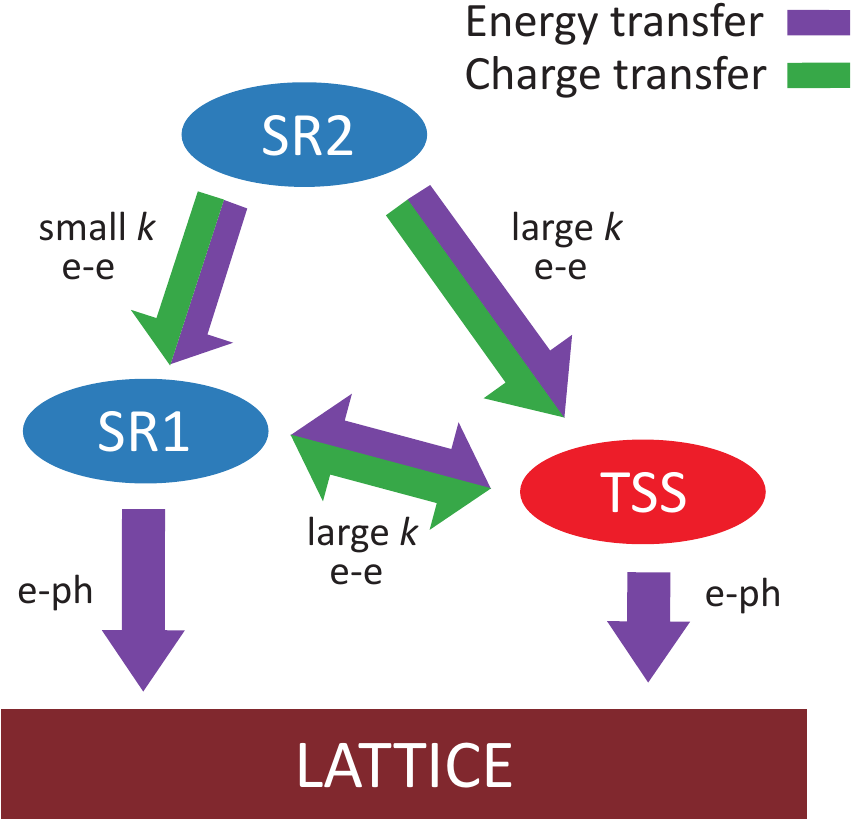} 
\caption{Schematic representation summarizing the relevant pathways for ultrafast charge and energy transfer among different bands in Bi$_2$Te$_3$, as well as the energy release into the lattice after optical excitation (see text). Blue and red colors emphasize the opposite spin textures of the surface-resonance states (SR$1$ and SR$2$) and the transient Dirac cone (TSS).}
\label{Figure5}
\end{figure}

In this respect, we emphasize that the relaxation times of the spin populations provide valuable information on the rates at which electrons in a given band release their energy and momentum, a process which is the result of a complex interplay between scattering probabilities associated with electron-electron and electron-phonon scatterings. Therefore, the observation that spin up and spin down electrons relax on similar time scales is fully compatible with the notion that electrons with opposite spins follow the same energy-, momentum- and spin-selection rules described by our theoretical calculations. This conclusion has unprecedented implications regarding the mechanisms underlying the electron dynamics. Hitherto, all time-resolved studies on TIs interpreted the ultrafast dynamics of their excited states solely on the basis of bulk-assisted electron-phonon scattering. However, our study reveals that electron-electron scattering processes on the surface -- that proceed on a time scale dictated by the spin texture of nonequlibrium states -- are the main driving force for electron relaxation. Thus, our work represents a paradigm shift with respect to previous interpretations of the ultrafast electron dynamics in TIs.

\vspace{0.06in}
{\bf E. Dynamical pathways of ultrafast charge and energy transfer}
\vspace{0.06in}

In Fig. 5 we show a simplified schematic representation that provides an overview of our present findings. The higher-energy surface resonances (SR2) relax through electron-electron scattering processes, and the overall transfer of charge and energy occurs through spin-dependent scattering into the lower-energy states (SR1 and TSS). The alternating spin textures of the different bands influences the possible electron transitions, since it establishes whether small (large) momentum transfers are allowed or forbidden, meaning that large momentum transfers are ultimately required for the relaxation of SR2 states. The time scale of these processes strongly influences the characteristic relaxation times of spin up and spin down electrons, triggering the subsequent dynamics, which proceeds on a slower time scale once electrons from the SR2 states have decayed into the lower-energy bands. Similarly, electron-electron scatterings lead to a continuous exchange of charge and energy between the TSS and SR1 bands, which in consequence behave as two dynamically-locked electron populations. The dynamics of such process is again altered by the existence of spin-forbidden transitions.
Subsequently, the energy of the electronic system is released into the lattice through electron-phonon scatterings, which govern the relaxation of the TSS and SR1 populations in the immediate vicinity of the Fermi level. 

Our findings reveal the crucial impact of spin-dependent transitions on the recombination processes of excited electrons on the surface of Bi$_2$Te$_3$, so that the complex physical picture underlying the ultrafast electron dynamics in this system cannot be understood without involving the electron spin. Therefore, we expect similar effects at play in the spin dynamics of other TI surfaces or systems where spin-orbit interactions are important, offering the possibility of manipulating spin degrees of freedom to optimize carrier lifetimes in future devices. Moreover, such manipulation would be accompanied by a full control of the device surface conductivity, as it would allow to modulate the relative importance of high-$k$ transfer transitions strongly contributing to the resistivity.

\vspace{0.1in}
\centerline{\bf IV. SUMMARY AND OUTLOOK}
\vspace{0.06in}

In conclusion, by the simultaneous energy, momentum, spin and sub-ps time resolution of our experiments in combination with unprecedentedly detailed calculations of the dynamics, we have unambiguously identified the scattering mechanisms underlying the ultrafast relaxation of individual bands in the prototypical topological insulator Bi$_2$Te$_3$. We have observed a transient population of spin-polarized Dirac fermions coexisting with two surface-resonance states above the Fermi level that relax through an avalanche of hot electrons within about $\sim$2 ps. One surface resonance has been found at high energies inside a bulk-projected band gap, the other one at low energies just above the border of the main bulk-band gap. We have observed that both surface resonances exhibit a nonequilibrium spin polarization that is reversed with respect to that of the transient Dirac bands, in agreement with our one-step model photoemission calculations. Their alternating spin polarization as a function of electron wave vector reveals that trivial and topological pure spin currents coexist on fs time scales. Finally, we have examined the momentum and energy dependence of the tr-ARPES intensities as well as the charateristic relaxation times of spin up and spin down electrons within different bands, and found excellent agreement with our dynamical calculations. Following this approach, we have been able to disentangle the contributions from electron-electron and electron-phonon scatterings quantitatively, and revealed the strong influence of spin-forbidden transitions and high-$k$ scattering in the decay pathways and relaxation time scales of excited electrons. 

The fundamental processes involving the electron spin revealed by our theoretical and experimental results are expected to have broad impact in spintronic applications, as they offer a rich playground for engineering spin textures to control not only the conductivity but also the relevant time scales of laser-induced pure spin currents and spin-polarized electrical currents in a  variety of different materials. It would be interesting to perform similar experiments as the one reported here in systems where the tunability of spin-forbidden transitions could be achieved, for example, via the Rashba effect, allowing to systematically control the lifetime of spin-polarized carriers. 

{\it Note added.} Recently, we became aware of a related time-resolved study on
Bi$_2$Se$_3$ \cite{Lanzara-NatComm-2016}.

\vspace{0.06in}
\centerline{\bf Acknowledgements}
\vspace{0.06in}

Financial support from the Deutsche Forschungsgemeinschaft (DFG Grant No. SPP 1666) is gratefully acknowledged. M. K. and A. V. acknowledge financial support from the Impuls-und Vernetzungsfonds der Helmholtz-Gemeinschaft (Grant No. HRJRG-408). J. M. is supported by the CENTEM project CZ.1.05/2.1.00/03.0088, co-funded by the ERDF as part of the Ministry of Education, Youth and Sports OP RDI programme. J. M., H. E. and K. H. acknowledge support from the DFG research unit FOR 1346. M. B. and K. H. acknowledge support from the Austrian Science Fund (FWF) through Lise Meitner grant M1925-N28, as well as from the European Research Council under the European Union's Seventh Framework Program (FP/2007-2013)/ERC grant agreement no. 306447. This work was partially supported by BMBF (05K10CBA).

J.S.-B., M.K. and E.G performed the experiments with assistance
from A.V., A.R., and O.K.; L.Y. provided bulk-single crystals
and performed sample characterization; M.B., J.B., J.M., H.E. and K. H.
carried out calculations; J.S.-B. and M. B. performed
data analysis, figure and draft planning; J.S.-B. and M.B. wrote the
manuscript with input from all authors; J.S.-B. was
responsible for the conception and the overall direction.



\begin{thebibliography}{20} 
  

\bibitem{Hasan-RMP-2010}
M. Z. Hasan and C. L. Kane, {\it Colloquium: Topological insulators}, Rev. Mod. Phys. {\bf 82}, 3045 (2010).

\bibitem{Moore-Nature-2010}
J. E. Moore, {\it The birth of topological insulators}, Nature {\bf 464}, 194–198 (2010).

\bibitem{Fu-PRL-2007}
L. Fu, C. L. Kane, and E. J. Mele, {\it Topological Insulators in Three Dimensions}, Phys. Rev. Lett. {\bf 98}, 106803 (2007).

\bibitem{Pesin-NatMat-2012}
D. Pesin and A. H. MacDonald, {\it Spintronics and pseudospintronics in graphene and topological insulators}, Nature Mater. {\bf 11}, 409 (2012).

\bibitem{Gedik-2012-NatNanotech-photocurrents}
J. W. McIver, D. Hsieh, H. Steinberg, P. Jarillo-Herrero and N. Gedik, {\it Control over topological insulator photocurrents with light polarization},  Nature Nanotech. {\bf 7}, 96 (2012).

\bibitem{Kastl-NatComm-2015}
C. Kastl, C. Karnetzky, H. Karl and A. W. Holleitner, {\it Ultrafast helicity control of surface currents in topological insulators with near-unity fidelity}, Nat. Commun. {\bf 6}, 6617 (2015).

\bibitem{Dankert-NanoLett-2015}
A. Dankert, J. G., M. V. Kamalakar, S. Charpentier, and S. P. Dash, {\it Room Temperature Electrical Detection of Spin Polarized Currents in Topological Insulators}, Nano Lett. {\bf 15}, 7976 (2015) 

\bibitem{Hsieh-Science-2009}
D. Hsieh, Y. Xia, L. Wray, D. Qian, A. Pal, J. H. Dil, J. Osterwalder, F. Meier, G. Bihlmayer, C. L. Kane, Y. S. Hor, R. J. Cava and M. Z. Hasan, {\it Observation of Unconventional Quantum Spin Textures in Topological Insulators}, Science {\bf 323}, 919 (2009).

\bibitem{Hsieh-Nature-2009}
D. Hsieh, Y. Xia, D. Qian, L. Wray, J. H. Dil, F. Meier, J. Osterwalder, L. Patthey, J. G. Checkelsky, N. P. Ong, A. V. Fedorov, H. Lin, A. Bansil, D. Grauer, Y. S. Hor, R. J. Cava, and M. Z. Hasan, {\it A tunable topological insulator in the spin helical Dirac transport regime}, Nature {\bf 460}, 1101 (2009).    

\bibitem{Jozwiak-PRB-2011}
Ch. Jozwiak , Y. L. Chen, A. V. Fedorov, J. G. Analytis, C. R. Rotundu, A. K. Schmid, J. D. Denlinger, Y.-D. Chuang, D.-H. Lee, I R. Fisher, R J. Birgeneau, Z.-X. Shen, Z. Hussain, and A. Lanzara, {\it Widespread spin polarization effects in photoemission from topological insulators}, Phys. Rev. B {\bf 84}, 165113 (2011).

\bibitem{Pan-PRB-2013}
Z.-H. Pan, E. Vescovo, A. V. Fedorov, G. D. Gu, and T. Valla, {\it Persistent coherence and spin polarization of topological surface states on topological insulators}, Phys. Rev. B {\bf 88}, 041101(R) (2013).

\bibitem{Sanchez-Barriga-PRX-2014}
J. S\'anchez-Barriga, A. Varykhalov, J. Braun, S.-Y. Xu, N. Alidoust, O.  Kornilov, J. Min\'ar, K. Hummer, G. Springholz, G. Bauer, R. Schumann, L. V. Yashina, H. Ebert, M. Z. Hasan, and O. Rader, {\it Photoemission of Bi$_{2}$Se$_{3}$ with Circularly Polarized Light: Probe of Spin Polarization or means for Spin Manipulation?}, Phys. Rev. X {\bf 4}, 011046 (2014).

\bibitem{Ren-PRB-2011}
Z. Ren, A. A. Taskin, S. Sasaki, K. Segawa, and Y. Ando, {\it Optimizing Bi$_{2-x}$Sb$_x$Te$_{3-y}$Se$_y$ solid solutions to approach the intrinsic topological insulator regime}, Phys. Rev. B {\bf 84}, 165311 (2011).

\bibitem{Arakane-NatComm-2012}
T. Arakane, T. Sato, S. Souma, K. Kosaka, K. Nakayama, M. Komatsu,	T. Takahashi, Z. Ren, K. Segawa, and Y. Ando, {\it Tunable Dirac cone in the topological insulator Bi$_{2-x}$Sb$_x$Te$_{3-y}$Se$_y$}, Nat. Commun. {\bf 3}, 636 (2012). 

\bibitem{Kuroda-PRB-2015}
K. Kuroda, G. Eguchi, K. Shirai, M. Shiraishi, M. Ye, K. Miyamoto, T. Okuda, S. Ueda, M. Arita, H. Namatame, M. Taniguchi, Y. Ueda, and A. Kimura, {\it Tunable spin current due to bulk insulating property in the topological insulator Tl$_{1-x}$Bi$_{1+x}$Se$_{2-\delta}$}, Phys. Rev. B {\bf 91}, 205306 (2015).


\bibitem{Gedik-PRL-2011-Kerr}
D. Hsieh, F. Mahmood, J. W. McIver, D. R. Gardner, Y. S. Lee, and N. Gedik, {\it Selective Probing of Photoinduced Charge and Spin Dynamics in the Bulk and Surface of a Topological Insulator}, Phys. Rev. Lett. {\bf 107}, 077401 (2011).

\bibitem{Sobota-PRL-2012-Bulk-Reservoir}
J. A. Sobota, S. Yang, J. G. Analytis, Y. L. Chen, I. R. Fisher, P. S. Kirchmann, and Z.-X. Shen, {\it Ultrafast Optical Excitation of a Persistent Surface-State Population in the Topological Insulator Bi$_2$Se$_3$}, Phys. Rev. Lett. {\bf 108}, 117403 (2012).

\bibitem{Hajlaoui-2012-Nanolett-bulk}
M. Hajlaoui, E. Papalazarou, J. Mauchain, G. Lantz, N. Moisan, D. Boschetto, Z. Jiang, I. Miotkowski, Y. P. Chen, A. Taleb-Ibrahimi, L. Perfetti, and M. Marsi, {\it Ultrafast Surface Carrier Dynamics in the Topological Insulator Bi$_2$Te$_3$}, Nano Lett. {\bf 12} , 3532 (2012).

\bibitem{Gedik-2012-PRL-phonons}
Y. H. Wang, D. Hsieh, E. J. Sie, H. Steinberg, D. R. Gardner, Y. S. Lee, P. Jarillo-Herrero, and N. Gedik, {\it Measurement of Intrinsic Dirac Fermion Cooling on the Surface of the Topological Insulator Bi$_2$Se$_3$ Using Time-Resolved and Angle-Resolved Photoemission Spectroscopy}, Phys. Rev. Lett. {\bf 109}, 127401 (2012).

\bibitem{Crepaldi-2012-PRB-phonons}
A. Crepaldi, B. Ressel, F. Cilento, M. Zacchigna, C. Grazioli, H. Berger, Ph. Bugnon, K. Kern, M. Grioni, and F. Parmigiani, {\it Ultrafast photodoping and effective Fermi-Dirac distribution of the Dirac particles in Bi$_2$Se$_3$}, Phys. Rev. B {\bf 86}, 205133 (2012). 

\bibitem{Luo-NanoLett-2013-increased-e-ph}
C. W. Luo, H. J. Wang, S. A. Ku, H.-J. Chen, T. T. Yeh, J.-Y. Lin, K. H. Wu, J. Y. Juang, B. L. Young, T. Kobayashi, C.-M. Cheng, C.-H. Chen, K.-D. Tsuei, R. Sankar, F. C. Chou, K. A. Kokh, O. E. Tereshchenko, E. V. Chulkov, Yu. M. Andreev, and G. D. Gu, {\it Snapshots of Dirac Fermions near the Dirac Point in Topological Insulators}, Nano Lett. {\bf 13}, 5797 (2013). 

\bibitem{Hajlaoui-NatComm-2014-e-h-asym}
M. Hajlaoui, E. Papalazarou, J. Mauchain, L. Perfetti, A. Taleb-Ibrahimi, F. Navarin, M. Monteverde, P. Auban-Senzier, C. R. Pasquier, N. Moisan, D. Boschetto, M. Neupane, M. Z. Hasan, T. Durakiewicz, Z. Jiang, Y. Xu, I. Miotkowski, Y. P. Chen, S. Jia, H. W. Ji, R. J. Cava, and M. Marsi, {\it Tuning a Schottky barrier in a photoexcited topological insulator with transient Dirac cone electron-hole asymmetry}, Nat. Commun. {\bf 5}, 3003 (2014).

\bibitem{Reimann-2014-PRB-phonons}
J. Reimann, J. G\"udde, K. Kuroda, E. V. Chulkov, and U. H\"ofer, {\it Spectroscopy and dynamics of unoccupied electronic states of the topological insulators $Sb_{2}Te_{3}$ and $Sb_{2}Te_{2}S$}, Phys. Rev. B {\bf 90}, 081106(R) (2014). 

\bibitem{Sobota-PRL-2014-Oscillations}
J. A. Sobota,  S.-L. Yang, D. Leuenberger, A. F. Kemper, J. G. Analytis, I. R. Fisher, P. S. Kirchmann, T. P. Devereaux, and Z.-X. Shen, {\it Distinguishing Bulk and Surface Electron-Phonon Coupling in the Topological Insulator Bi$_2$Se$_3$ Using Time-Resolved Photoemission Spectroscopy}, Phys. Rev. Lett. {\bf 113}, 157401 (2014).


\bibitem{Gedik-2013-Science-Floquet}
Y. H. Wang, H. Steinberg, P. Jarillo-Herrero, and N. Gedik, {\it Observation of Floquet-Bloch States on the Surface of a Topological Insulator}, Science {\bf 342}, 453 (2013).

\bibitem{Gedik-2016-NaturePhys-Floquet}
F. Mahmood, C.-K. Chan, Z. Alpichshev, D. Gardner, Y. Lee, P. A. Lee, and N. Gedik, {\it Selective scattering between Floquet–Bloch and Volkov states in a topological insulator}, Nat. Phys. {\bf 12}, 306 (2016).

\bibitem{Kuroda-PRL-2016}
K. Kuroda, J. Reimann, J. G\"udde, and U. H\"ofer, {\it Generation of Transient Photocurrents in the Topological Surface State of Sb$_2$Te$_3$ by Direct Optical Excitation with Midinfrared Pulses}, Phys. Rev. Lett. {\bf 116}, 076801 (2016)

\bibitem{Dahlhaus-PRL-2016}
J. P. Dahlhaus, Benjamin M. Fregoso, and Joel E. Moore, {\it Magnetization Signatures of Light-Induced Quantum Hall Edge States}, Phys. Rev. Lett. {\bf 114}, 246802 (2015). 


\bibitem{Cacho-Surface-resonances-PRL-2014}
C. Cacho, A. Crepaldi, M. Battiato, J. Braun, F. Cilento, M. Zacchigna, M. C. Richter, O. Heckmann, E. Springate, Y. Liu, S. S. Dhesi, H. Berger, Ph. Bugnon, K. Held, M. Grioni, H. Ebert, K. Hricovini, J. Min\'ar, and F. Parmigiani, {\it Momentum-Resolved Spin Dynamics of Bulk and Surface Excited States in the Topological Insulator Bi$_2$Se$_3$}, Phys. Rev. Lett. {\bf 114}, 097401 (2015).

\bibitem{Sanchez-Barriga-PRB-2016}
J. S\'anchez-Barriga, E. Golias, A. Varykhalov, J. Braun, L. V. Yashina, R. Schumann, J. Minár, H. Ebert, O. Kornilov, and O. Rader, {\it Ultrafast spin-polarization control of Dirac fermions in topological insulators}, Phys. Rev. B {\bf 93}, 155426 (2016).

\bibitem{Boschini-SciRep-2016}
F. Boschini, M. Mansurova, G. Mussler, J. Kampmeier, D. Grützmacher, L. Braun, F. Katmis, J. S. Moodera, C. Dallera, E. Carpene, C. Franz, M. Czerner, C. Heiliger, T. Kampfrath, and M. Münzenberg, {\it Coherent ultrafast spin-dynamics probed in three dimensional topological insulators}, Sci. Rep. {\bf 5}, (2015).


\bibitem{Pascual-PRL-2004}
J. I. Pascual, G. Bihlmayer, Yu. M. Koroteev, H.-P. Rust, G. Ceballos, M. Hansmann, K. Horn, E. V. Chulkov, S. Blügel, P. M. Echenique, and Ph. Hofmann, {\it Role of Spin in Quasiparticle Interference}, Phys. Rev. Lett. {\bf 93}, 196802 (2004).

\bibitem{Roushan-Nature-2009}
P. Roushan, J. Seo, C. V. Parker, Y. S. Hor, D. Hsieh, D. Qian, A. Richardella, M. Z. Hasan, R. J. Cava and A. Yazdani, {\it Topological Surface States Protected From Backscattering by Chiral Spin Texture}, Nature {\bf 460}, 1106 (2009).


\bibitem{Wang-PRL-2016}
M.C. Wang, S. Qiao, Z. Jiang, S.N. Luo, and J. Qi, {\it Unraveling Photoinduced Spin Dynamics in the Topological Insulator Bi$_2$Se$_3$}, Phys. Rev. Lett. {\bf 116}, 036601 (2016).

\bibitem{Sim-PRB-2014}
S. Sim, M. Brahlek, N. Koirala, S. Cha, S. Oh, and H. Choi, {\it 
Ultrafast terahertz dynamics of hot Dirac-electron surface scattering in the topological insulator Bi$_2$Se$_3$}, Phys. Rev. B {\bf 89}, 165137 (2014).


\bibitem{Hopkinson-CPC-1980}
J. F. L. Hopkinson, J. B. Pendry, and D. J. Titterington, {\it Calculation of photoemission spectra for surfaces of solids}, Comput. Phys. Commun. {\bf 5}, 599 (1980).

\bibitem{Braun-theory-96} 
J. Braun, {\it The theory of angle-resolved ultraviolet photoemission and its applications to ordered materials}, Rep. Prog. Phys. {\bf 59}, 1267 (1996).

\bibitem{Ebert-SPRKKR-2011}
H. Ebert, D. K\"odderitzsch, and J. Min\'ar, {\it Calculating condensed matter properties using the KKR-Green's function method–recent developments and applications}, Rep. Prog. Phys. {\bf 74}, 096501 (2011).

\bibitem{Ebert-SPRKKR-2012}
H. Ebert, {\it The Munich SPR-KKR program package, version 6.3}, http://olymp.cup.uni-muenchen.de/ak/ebert/SPRKKR (2012).


\bibitem{Fatti-PRB-2000}
N. Del Fatti, C. Voisin, M. Achermann, S. Tzortzakis, D. Christofilos, and F. Vall\'ee, {\it Nonequilibrium electron dynamics in noble metals}, Phys. Rev. B {\bf 61}, 16956 (2000).

\bibitem{Rethfeld-PRB-2002}
B. Rethfeld, A. Kaiser, M. Vicanek, and G. Simon, {\it Ultrafast dynamics of nonequilibrium electrons in metals under femtosecond laser irradiation}, Phys. Rev. B {\bf 65}, 214303 (2002).

\bibitem{Mueller-PRB-2013}
B. Y. Mueller and B. Rethfeld, {\it Relaxation dynamics in laser-excited metals under nonequilibrium conditions}, Phys. Rev. B {\bf 87}, 035139 (2013).

\bibitem{Supplemental-Material}
See Supplemental Material at [URL] for more details on theoretical aspects of the calculations.

\bibitem{Echenique-JPC-1978}
P. M. Echenique and J. B. Pendry, {\it The existence and detection of Rydberg states at surfaces}, J. Phys. C {\bf 11}, 2065 (1978).

\bibitem{McRae-RMP-1979}
E. G. McRae, {\it Electronic surface resonances of crystals}, Rev. Mod. Phys. {\bf 51}, 541 (1979).

\bibitem{Zhu-PRL-2013}
Z. H. Zhu, C. N. .Veenstra, G. Levy, A. Ubaldini, P. Syers, N. P. Butch, J. Paglione, M. W. Haverkort, I. S. Elfimov, and A. Damascelli, {\it Layer-By-Layer Entangled Spin-Orbital Texture of the Topological Surface State in Bi$_2$Se$_3$}, Phys. Rev. Lett. {\bf 110}, 216401 (2013).

\bibitem{Optical-Orientation-1984}
F. Meier and B. Zakharchenya, {\it Optical Orientation} (Elsevier, Amsterdam, 1984). 

\bibitem{Foester-PRB-2015} 
T. F\"orster, P. Kr\"uger, and M. Rohlfing, {\it Two-dimensional topological phases and electronic spectrum of Bi$_2$Se$_3$ thin films from GW calculations}, Phys. Rev. B {\bf 92}, 201404(R) (2015).

\bibitem{Hsieh-PRL-2009} 
D. Hsieh, Y. Xia, D. Qian, L. Wray, F. Meier, J. H. Dil, J. Osterwalder, L. Patthey, A. V. Fedorov, H. Lin, A. Bansil, D. Grauer, Y. S. Hor, R. J. Cava, and M. Z. Hasan, {\it Observation of Time-Reversal-Protected Single-Dirac-Cone Topological-Insulator States in Bi$_2$Te$_3$ and Sb$_2$Te$_3$}, Phys. Rev. Lett. {\bf 103}, 146401 (2009).

\bibitem{Guede-PRB-2015}
J. G\"udde, M. Rohleder, T. Meier, S. W. Koch, U. H\"ofer, {\it Time-Resolved Investigation of Coherently Controlled Electric Currents at a Metal Surface}, Science {\bf 318}, 1287 (2007).

\bibitem{Hellman-PRB-2015}
O. Hellman and D. A. Broido, {\it Phonon thermal transport in Bi$_2$Te$_3$ from first principles}, Phys. Rev. B {\bf 90}, 134309 (2014).

\bibitem{Lanzara-NatComm-2016}
C. Jozwiak, J. A. Sobota, K. Gotlieb, A. F. Kemper, C. R. Rotundu,
R. J. Birgeneau, Z. Hussain, D.-H. Lee, Z.-X. Shen, and A. Lanzara, {\it Spin-polarized surface resonances accompanying
topological surface state formation}, Nat. Commun. {\bf 7}, 13143 (2016).

\end{thebibliography}
\end{document}